# Reconstruction, with tunable sparsity levels, of shear-wave velocity profiles from surface wave data

Giulio Vignoli[1,2], Julien Guillemoteau[3], Jeniffer Barreto[1], and Matteo Rossi[4]

[1]*University of Cagliari, Cagliari, 09123 Italy*
[2]*Geological Survey of Denmark and Greenland, Aarhus, 8000 Denmark*
[3]*University of Potsdam, Potsdam, 14469 Germany*
[4]*Lund University, SE-221 00 Lund, Sweden*

*Corresponding author: G. Vignoli (e-mail: gvignoli@unica.it).*

**SUMMARY**

The analysis of surface wave dispersion curves is a way to infer the vertical distribution of shear-wave velocity. The range of applicability is extremely wide: going, for example, from seismological studies to geotechnical characterizations and exploration geophysics. However, the inversion of the dispersion curves is severely ill-posed and only limited efforts have been put in the development of effective regularization strategies. In particular, relatively simple smoothing regularization terms are commonly used, even when this is in contrast with the expected features of the investigated targets. To tackle this problem, stochastic approaches can be utilized, but they are too computationally expensive to be practical, at least, in case of large surveys. Instead, within a deterministic framework, we evaluate the applicability of a regularizer capable of providing reconstructions characterized by tunable levels of sparsity. This adjustable stabilizer is based on the minimum support regularization, applied before on other kinds of geophysical measurements, but never on surface wave data. We demonstrate the effectiveness of this stabilizer on: i) two benchmark – publicly available – datasets at crustal and near-surface scales; ii) an experimental dataset collected on a well-characterized site. In addition, we discuss a possible strategy for the estimation of the depth of investigation. This strategy relies on the integrated sensitivity kernel used for the inversion and calculated for each individual propagation mode. Moreover, we discuss the reliability, and possible caveats, of the direct interpretation of this particular estimation of the depth of investigation, especially in the presence of sharp boundary reconstructions.

# 1 INTRODUCTION

Surface wave data are routinely used for the reconstruction of the elastic properties of the subsoil (in particular, the shear-wave velocity) at depths ranging from near-surface (Xia et al. 1999; Laake & Strobbia 2012; Park et al. 2017) to crustal scales (Dorman & Ewing 1962; Beucler et al. 2003; Ekström 2011; Ars et al. 2019). These data, usually, consist of dispersion curves extracted from the seismograms through, for example, integral transforms (Vignoli & Cassiani 2010; Vignoli et al. 2011; Pan et al. 2019). Dispersion curves describe the dependence of the surface wave (group and/or phase) velocities with respect to the frequencies. Dispersion curves are, in turn, inverted by using 1D forward modelling algorithms (Park et al. 1999). The inferred shear-wave velocity vertical profiles are, then, stitched together (Vignoli et al. 2016) or even jointly inverted by enforcing some level of spatial coherency between adjacent 1D profiles (Wisén & Christiansen, 2005) in order to generate (pseudo-)2D/3D sections.

The inversion of dispersion curves is a typical ill-posed problem and the associated ambiguity is tackled with different kinds of regularizations. Many of these strategies are based on smoothing regularizations accomplished via the penalization of the L2-norm of a roughening operator applied to the model parameter vector, no matter if relying on single soundings (Vignoli et al. 2012a), on neighboring dispersion curves (Bergamo et al. 2016), on ancillary information (Cercato 2008) or additional geophysical observations (Wu et al. 2018).

Alternative inversion methodologies are based on stochastic approaches; e.g. Wathelet et al. (2004) Dal Moro et al. (2007), Maraschini & Foti (2010). In general, these latter strategies have several advantages compared to the deterministic ones. For example, they are capable to explore more effectively the model parameter space and, by doing so, to naturally provide a more comprehensive estimation of the uncertainty of the result. However, their main drawback is still the computational cost. Thus, at least in the (more and more common) case of large datasets (Strobbia et al. 2011), it is highly desirable to have a fast inversion tool that, at the same time, is flexible enough to incorporate a large class of different kinds of prior information.

In this perspective, in the present study, we evaluate a modification of a "standard" L2-norm approach that, instead of selecting the smoothest possible solution, promotes "sparse" reconstructions (of course, all equally compatible with the observations). Here, "sparse" should be intended in a broad sense, as the retrieved solutions can actually be quite smooth depending on the choice of the tuning parameter. The main difference between the standard L2-norm regularization and the proposed sparse strategy is that, whereas the first is favoring solutions with the smallest spatial variations of the shear-wave velocities, the latter is selecting the model characterized by the smallest number of model parameters in which the velocity variations occur (regardless of the actual amount of the variation). The proposed deterministic inversion approach is based on the minimum support regularization. Recently, other strategies, making use of L1-norm regularizations have been applied to dispersion curve data for the reconstruction of blocky solutions (Haney & Qu 2010; Esfahani et al. 2020). However, here, we would like to highlight the flexibility of the minimum support approach capable of retrieving shear-wave velocity profiles with an adjustable level of sparsity. We first demonstrate the performances of the novel tunable approach via synthetic datasets available in the literature. In particular, we discuss the effects of the choice of the free parameter characterizing the proposed regularization and how, in principle, it can be used to explore the model parameter space. The two considered examples are at very different scales and deal with the inversion of different numbers of propagation modes. In addition, we verify our conclusions by applying the proposed inversion strategy to an experimental dataset collected for geotechnical investigations on a well-characterized site.

In the literature, there are several examples about the applications of this type of regularization on different geophysical measurements; just to mention few examples: travel-times (Vignoli & Zanzi 2005; Zhdanov et al. 2006; Ajo-Franklin et al. 2007), electrical resistivity tomography (Pagliara & Vignoli 2006; Fiandaca et al. 2015), electromagnetic observations (Zhdanov & Tolstaya 2004; Ley-Cooper et al. 2015; Vignoli et al. 2017; Dragonetti et al. 2018), and gravimetric measurement (Last & Kubik 1983). However, the present research is the first attempt of dealing with seismic surface waves. In addition, to facilitate the interpretation of the results, we further develop the analysis in Haney & Tsai (2017) and describe a possible approach for the assessment of the depth of investigation associated with each mode of propagation.

## 2 INVERSION SCHEME

In studying the surface wave propagation in layered systems, a crucial problem is the determination of the dispersion curves given the elastic properties of the subsurface. This problem is highly nonlinear and can be addressed in several ways (Thomson 1950; Haskell 1953; Lysmer 1970). Nevertheless, independently from the forward modelling formulation used, the corresponding inverse problem can be solved via an iterative process making use of local linearizations of the original problem (Zhdanov 2002).

As discussed in detail in Aki & Richards (2002) and Haney & Tsai (2017), the linear relationship between the relative perturbations in the phase velocity $c$ and in shear-wave velocity $\beta$ can be expressed

as

$$\frac{\delta \mathbf{c}}{\mathbf{c}} = \mathbf{K}\frac{\delta \boldsymbol{\beta}}{\boldsymbol{\beta}}, \qquad (1)$$

where **K** is the shear-wave velocity kernel, and **c** and **β** are two vectors, as the first is evaluated over many frequencies (and, possibly, several modes of propagation), whereas the latter consists of the velocities of the numerous layers used for discretizing the subsurface.

Under the frequent assumptions of a constant density model and a fix Poisson's ratio, **K** is simply twice the sum of the kernels for the shear modulus and the Lamé's first parameter. These common assumptions are justified, for example, by parametric analyses confirming that phase velocities are mainly affected by the shear-wave velocity distributions (Xia et al. 1999) and that the mass density is characterized by limited ranges and impacts (Ivanov et al. 2016). Consistently, Poisson's ratio is, in general, considered known. A homogeneous Poisson's ratio is reasonable, especially for deep investigations (Xing et al. 2016, Esfahani et al. 2020), whereas, for shallower explorations, the effect of the saturation on compressional-wave velocity (and, in turn, on Poisson's ratio) can be significant; in these cases, for example, the depth of the water-table should be taken into account for the proper assessment of the Poisson's ratio variability (Foti & Strobbia 2002). The examples discussed in the present paper are all performed accordingly to these two abovementioned hypotheses concerning the homogeneous mass density and constant Poisson's ratio.

From (1), it is easy to see that the linear Jacobian operator, locally mapping the shear-velocity onto the phase velocity, is

$$\mathbf{J} = diag(\mathbf{c})\,\mathbf{K}\,diag(\boldsymbol{\beta})^{-1}, \qquad (2)$$

in which $\text{diag}(\mathbf{c})$ is the square diagonal matrix with the nonzero entries equal to the elements of **c**, and, similarly, $\text{diag}(\boldsymbol{\beta})^{-1}$ is the inverse of the diagonal matrix having as diagonal the vector **β**.

Thus, within the inversion scheme described, for example, in Tarantola & Valette (1982), the model **β** can be updated accordingly to the following recursive expression:

$$\begin{bmatrix} \mathbf{W}_d \\ \mathbf{0} \end{bmatrix} \left(\mathbf{c}_d - f(\boldsymbol{\beta}_{n-1}) + \mathbf{J}(\boldsymbol{\beta}_{n-1} - \boldsymbol{\beta}_0)\right) = \begin{bmatrix} \mathbf{W}_d\,\mathbf{J} \\ \mathbf{W}_m \end{bmatrix} (\boldsymbol{\beta}_n - \boldsymbol{\beta}_0), \qquad (3)$$

where $\mathbf{c}_d$ represents the observed dataset, $f$ is the nonlinear forward modelling operator (whose Fréchet derivative calculated in $\boldsymbol{\beta}_{n-1}$ is **J**), $\boldsymbol{\beta}_0$ is the reference (and, in our case, also, the initial) model, and $\mathbf{W}_d$ and $\mathbf{W}_m$ are the weighting matrices for the data and the model vectors, respectively. Clearly, the goal is to find, after $n$ iterations, the shear-wave velocity update $\Delta \boldsymbol{\beta}_n = (\boldsymbol{\beta}_n - \boldsymbol{\beta}_0)$ minimizing the norm of distance between the two sides of the augmented system of equations in (3), i.e. $\left\| \mathbf{W}_d \left(\mathbf{c}_d - f(\boldsymbol{\beta}_n)\right) \right\|_{L_2}^2 + \left\| \mathbf{W}_m \Delta \boldsymbol{\beta}_n \right\|_{L_2}^2$. In this paper, the number of iterations $n$ is dynamically

defined as the number of iterations necessary to reach a stationary point characterized by no significant variations in the value of the objective functional resulting from (3).

$\mathbf{W}_d$ is related to the data covariance $\mathbf{C}$, and, in our case, by assuming mutually independent data, it can be considered equal to

$$\mathbf{W}_d = \text{diag}(\boldsymbol{\sigma}_d)^{-1} = \mathbf{C}^{-1/2}, \tag{4}$$

where the $i$-th component of the vector $\boldsymbol{\sigma}_d$ is the standard deviation of the phase velocity associated with the $i$-th frequency.

If we select $\mathbf{W}_m$ in (3) equal to $\lambda \mathbf{L}$ – with (i) $\lambda$ being the Tikhonov parameter controlling the relative importance of the regularization term with respect of the data misfit, and (ii) $\mathbf{L}$ a discrete approximation of the spatial derivative – then, the system of equations (3) provides the standard Minimum Gradient Norm (MGN) solution. In case of a homogeneous reference model $\boldsymbol{\beta}_0$ (i.e. in the case of a spatial variation of the reference model $\boldsymbol{\beta}_0$ being zero everywhere), the MGN solution is characterized by the minimum velocity variations between the adjacent layers of the subsurface parameterization (strictly speaking, the MGN stabilizer guarantees the uniqueness of the solution only if the forward problem is linear. Clearly, this is not the case when dealing with dispersion curve inversion. Thus, in this circumstance, the previous statement must be considered true only locally).

If, instead, $\mathbf{W}_m$ is chosen to be

$$\mathbf{W}_m = \lambda \, \text{diag}(\mathbf{W}_e) \, \mathbf{L}, \tag{5}$$

with

$$\mathbf{W}_e = ((\mathbf{L} \, \Delta \boldsymbol{\beta}_n)^2 + \varepsilon^2)^{-1/2}, \tag{6}$$

then, we are dealing with the Minimum Gradient Support (MGS) (Zhdanov et al. 2006). In the case of a homogeneous reference model, the MGS solution is characterized by the minimum number of parameters (i.e. layers) where a significant velocity variation between the adjacent layers occurs (again, strictly speaking, the same caveat concerning the uniqueness of the MGN solution is valid also for the MGS stabilizer). So, compared to the MGN, in the MGS, what matters is not the amount of the variation, but, rather, the number of parameters where a significant variation takes place (Vignoli et al. 2012b). A velocity variation contributes significantly to the MGS regularization term in (3) every time it is large compared to the threshold defined by the focusing parameter $\varepsilon$ (Vignoli et al. 2015). Thus, as it will be clear from the numerical examples in the following, a strategy for the selection of the optimal $\varepsilon$ value can consist in setting the focusing parameter value in the order of magnitude of the velocity changes we are interested to investigate. In principle, the focusing parameter $\varepsilon$ can be a vector and, in this way, depth-dependent sparsity of the solution can be enforced. For sake of simplicity, in the present research, we consider all the components of such a vector equal to $\varepsilon$; i.e. $\boldsymbol{\varepsilon} = \varepsilon\mathbf{1}$. In the described inversion scheme, the Tikhonov parameter $\lambda$ can be potentially chosen in many ways – e.g., through the L-curve strategy (Farquharson & Oldenburg 2004). Since the data uncertainty was accessible in the available test datasets, we select $\lambda$ in order to guarantee a chi-squared value $\chi^2 = (1/N_d)\|\mathbf{W}_d(\mathbf{c}_d - \boldsymbol{f}(\boldsymbol{\beta}_n))\|_{L_2}^2$ approximately equal 1 (with $N_d$ being the number of measurements) (Vignoli et al. 2012b).

It might be worth looking at the proposed tunable approach from a Bayesian perspective. In that framework, the goal is to study the posterior probability density function $p(\boldsymbol{\beta}|\mathbf{c}_d)$ measuring the probability of having the model $\boldsymbol{\beta}$ compatible with the measurements $\mathbf{c}_d$. Accordingly to the Bayes' theorem, $p(\boldsymbol{\beta}|\mathbf{c}_d)$ is proportional to: i) the prior probability density function for the model parameters $p(\boldsymbol{\beta})$, and the conditional probability density function $p(\mathbf{c}_d|\boldsymbol{\beta})$. The latter connects the measured data and the model parameters, and, in the specific case of a Gaussian noise distribution, it can be written as $p(\mathbf{c}_d|\boldsymbol{\beta}) = k_d \exp\left(-(\mathbf{c}_d - \boldsymbol{f}(\boldsymbol{\beta}))^T \mathbf{W}_d^T \mathbf{W}_d (\mathbf{c}_d - \boldsymbol{f}(\boldsymbol{\beta}))\right)$, where $k_d$ is merely a normalization factor. If also the model parameters are assumed to follow a Gaussian distribution, then, the prior information about the solution can be formalized as follows: $p(\boldsymbol{\beta}) = k_m \exp\left(-(\boldsymbol{\beta} - \boldsymbol{\beta}_0)^T \mathbf{W}_m^T \mathbf{W}_m (\boldsymbol{\beta} - \boldsymbol{\beta}_0)\right)$, in which: i) $k_m$ is another normalization factor and ii) the Gaussian is centered on the reference model $\boldsymbol{\beta}_0$. By comparing the posterior probability density function $p(\boldsymbol{\beta}|\mathbf{c}_d) \propto p(\mathbf{c}_d|\boldsymbol{\beta})p(\boldsymbol{\beta})$ and the objective functional $\|W_d(\mathbf{c}_d - \boldsymbol{f}(\boldsymbol{\beta}_n))\|_{L_2}^2 + \|W_m \Delta\boldsymbol{\beta}_n\|_{L_2}^2$ minimized by the system of equations in (3), it is evident that the proposed deterministic MGN algorithm is designed to search for the maximizer of the $p(\boldsymbol{\beta}|\mathbf{c}_d)$ via linearized steps with the assumption of a Gaussian distribution of the data noise and of the variation of the model. Along the same line of reasoning, the MGS can be seen as related to the assumption of a model variation distribution that is still Gaussian, but with a standard deviation $\mathbf{W}_e^{-1}$ that is, iteration by iteration, locally defined. In particular, such a standard deviation tends to the constant value $\varepsilon$, anywhere the associated variation of the model update $\Delta\boldsymbol{\beta}_n$ is small with respect to $\varepsilon$, whereas it is proportional to the variation of $\Delta\boldsymbol{\beta}_n$, whenever the model update itself is large compared to the threshold (defined by $\varepsilon$). That is why, during the inversion, large variations are all penalized (approximately) by the same amount, no matter how big they are (in fact, large variations are associated to proportionally large standard deviations, and their penalization is, in any case, close to 1), whereas small variations of the model update are penalized depending on their departure from 0 with respect to the $\varepsilon$ value.

Clearly, for the MGN stabilizer, the model covariance is assumed to be such that $\mathbf{C}_m^{-1} = \lambda^2 L^T L$,

whereas, for the MGS, the relation $\mathbf{C}_m^{-1} = \lambda^2 \mathbf{L}^T \mathbf{L}/(\Delta\boldsymbol{\beta}_n^T \mathbf{L}^T \mathbf{L}\Delta\boldsymbol{\beta}_n + \varepsilon^2)$ is valid.

Of course, the convenience of the deterministic local inversion algorithm, even if computationally cheaper than the global counterpart, comes with a price. In fact, compared with global approaches (e.g., Zunino et al. 2015), the proposed scheme starts from an initial guess, which is also the reference model $\boldsymbol{\beta}_0$ (i.e., our best guess), and searches in the direction obtained by the neighborhood derivative; in doing so, the local strategy leads to a minimum that is close to the initial guess and that, in principle, might not be the global optimum. However, it is important to highlight that, usually, the convergence to the global minimum cannot be claimed within a finite time even for the global techniques (Nelles 2013). Luckily, in the inversion of surface wave dispersion curves, a quite good (at least, homogeneous) starting model can be deduct from a direct observation of the data. In fact, the maximum phase velocity can be considered a reasonable estimation of the final maximum shear-wave velocity; therefore, generally, the proposed deterministic local approach should be able to perform quite well anytime proper starting models are selected based on a very easy preliminary analysis (calculation of the maximum) of the observed data.

Concerning the assumption of a diagonal data covariance $\mathbf{C}$, clearly, it is a modelling simplification and, most likely, the effectiveness of the proposed scheme can be improved by including, for example, an estimation of the error introduced by the 1D interpretation of the dispersion curves. Indeed, this additional piece of information might be incorporated into the covariance matrix (under the assumption that also the modelling error is Gaussian) by simply defining a new covariance matrix $\mathbf{C}'$ as the summation, $\mathbf{C}' = \mathbf{C} + \mathbf{C}_\Delta$, of the original term $\mathbf{C}$ of equation (4) and an assessment of the modelling error $\mathbf{C}_\Delta$ (Tarantola 2005). For example, the additional covariance $\mathbf{C}_\Delta$ can be calculated statistically by performing a number of accurate forward simulations taking into account the higher dimensionality of the real problem and comparing the associate responses with those of the 1D forward modelling used for the actual inversion (Hansen et al. 2014, Madsen & Hansen 2018, Zunino & Mosegaard 2019).

### 2.1 Depth of investigation

The Depth of Investigation (DOI) can be defined as the depth below which the data collected at the surface are not sensitive to the ground velocities. Hence, the DOI provides a rough estimation of the maximum depth that can be reliably investigated from the surface (given a specific frequency range, the number of considered modes, and, clearly, the physical properties of the medium).

A DOI assessment can be based on empirical relations relying on the longest wavelength in the data (Park & Carnevale 2010). Alternatively, the study of the variability of the solution as function of the starting model can provide good indications of the sensitivity of the data to the velocity values inferred at depth. So, for example, Foti et al. (2018) discusses the possibility of inverting the data with different initial velocity models and comparing the results to determine if they are data- or model-driven.

Similarly to Luo et al. 2007, the approach used here is based on the integrated sensitivity matrix (Zhdanov 2002). Hence, the DOI in the following examples is defined as the depth where, for each propagation mode, the normalized integrated sensitivity value drops by 70 dB.

The rationale behind this is that the variation $\delta[\mathbf{c}]_i = \delta[f(\boldsymbol{\beta})]_i$ – induced on the $i$-th phase velocity value by the perturbation $\delta[\boldsymbol{\beta}]_k$ of the $k$-th component of shear-velocity – is $\mathbf{J}_{ik}$ (if the linearized problem is considered). Thus, the overall influence of the perturbation of the $k$-th model parameter on the norm of the data vector is $[\mathbf{S}]_k = \sqrt{\sum_i (\delta[\mathbf{c}]_i)^2} = \left[\sqrt{\mathbf{J}^T \mathbf{J}}\right]_k$. Therefore, the various components of the integrated sensitivity vector $\mathbf{S}$ quantify the influence of a specific shear-velocity layer on the observed data. When $[\mathbf{S}]_k$ reduces significantly, the reconstruction should be no longer assumed data-driven. This is the depth where the DOI can be set. Hence, the DOI collapses the $\mathbf{S}$ information content in one number (actually, in one number for each data subset since we are possibly dealing with more than one propagation mode). Definitely, in the operative definition of DOI, the selection of the 70 dB threshold is quite arbitrary, and it has been decided a-posteriori based on some numerical tests. In any case, the same threshold has been applied across all the discussed inversions; so, at least, it can provide a consistent term of comparison.

## 3   COMPARISON OF THE INVERSION RESULTS

Following the approach in Haney & Tsai (2017), two synthetic datasets are inverted to demonstrate the capabilities of the two regularization strategies considered in this paper and defined by the two alternative choices for the model weight $\mathbf{W}_m$ in the previous section.

### 3.1 Crustal scale example

The first example concerns the inversion of the first two modes of propagation for a crustal scale problem. The frequency range of the data spans between 0.1 and 0.65 Hz. The shear-velocity model to be reconstructed (Fig. 1, solid blue line) is characterized by a ~2.2 km thick low-velocity zone, centered at around 3 km depth. Apart from this shallow velocity inversion, the shear-velocity profile shows an increasing velocity from 2.3 km/s to 3.5 km/s, with an intermediate step at 2.6 km/s. Accordingly to Haney & Tsai (2017), the data were contaminated with 2.5% of noise on both modes (Fig. 2, blue vertical bars).

The novel MGS ($\lambda = 5\ 10^{-2}$; $\varepsilon = 10^2$) reconstruction is showed in Fig. 1a and is supposed to be compared with the "standard" MGN ($\lambda = 5\ 10^{-8}$) result in Fig. 1b. Both these inversions have a $\chi^2$ value around 0.93. So, they are equally compatible with the data and in good agreement with the estimation of the noise level. Nevertheless, they show quite different features. In particular, despite the fact that both inferred models lack retrieving the shallow velocity inversion, this first MGS ($\lambda = 5\ 10^{-2}$; $\varepsilon = 10^2$) inversion can better identify the presence of sharp interface at around 4 km, and its velocities are very close to the true model down to 16 km (including the location of the velocity change at 16.3 km depth). This MGS ($\lambda = 5\ 10^{-2}$; $\varepsilon = 10^2$) inversion overestimates, by approximately 8%, the velocity of the last 3.5 km/s layer. This is not surprising as that layer lies in the depth interval between the two DOIs (Fig. 1a). In this case, the plot of the normalized integrated sensitivity (Fig. 3) demonstrates the potential risks connected with a too rigid interpretation of the DOI value: even if the sensitivity decay associated with the fundamental mode reaches the 70 dB threshold above the abrupt velocity change (in fact, the 1st DOI is set to ~15 km depth), the dramatic variation of the sensitivity actually occurs at the interface located at ~16 km; so, whereas the reconstruction above the abrupt (velocity/sensitivity) change can be considered reliable (even below the 1st DOI), the inferred velocity value for the last layer should be taken with more caution. On the other hand, the DOI for the MGN reconstruction is safer to be interpreted directly since, by definition, no sharp velocity (and in turn, sensitivity) changes are introduced (Fig. 4). Besides, this sensitivity analysis confirms that higher modes can be very valuable in bringing additional information at depth. The sensitivity decay is, indeed, less pronounced for the overtones, in particular, even below sharp transitions, where the fundamental mode sensitivity diminishes very rapidly.

Actually, the estimation of the originally provided noise might have been too conservative. In fact, by reducing the value of $\lambda$, we can get the result in Fig. 1c, in which the true model is effectively reconstructed (including the shallow velocity inversion) by the MGS ($\lambda = 5\ 10^{-3}$; $\varepsilon = 10^2$) algorithm. The corresponding $\chi^2$ is around 0.81. The only appreciable differences between the latest MGS result and the true model are: (i) the 9% underestimation of the depth of the deepest velocity variation, and (ii) an increasing velocity at the bottom of the profile. Both these differences occur below the sudden decrease of sensitivity (Fig. 5). Reaching the same level of data fitting (i.e. $\chi^2 \sim 0.81$) with the MGN regularization was not possible. This is in agreement with the fact that, through the MGN regularization, we try to enforce prior information in contrast with the actual blocky nature of the true model (Tarantola 2006).

Results similar to the MGS inversions in Fig. 1a and Fig. 1c can be probably obtained with a few layers parameterization. However, in practice: i) it is hard to know in advance the appropriate number of layers for the correct parameterization, and ii) the layer number might not be constant over a large survey with many adjacent acquisition locations (potentially requiring continuous supervised adjustments of the inversion settings). On the contrary, all the inversions discussed in this example are performed with layers with constant thickness (0.25 km). This provides a very flexible tool for the inversion of surface wave data in complex (blocky) geological settings, in which: i) the regularization parameter λ is selected by the discrepancy principle (in general, $\chi^2 \sim 1$), and ii) the focusing parameter ε is chosen accordingly to the expected sparsity level. In this respect, for this first experiment, we use $\varepsilon = 10^2$ since we are targeting shear-velocity variations of that order of magnitude; clearly, velocity variations of the order of few hundreds of meters per second can be considered significant (and penalized by the stabilizer), not only as they characterize the true model, but, in general, because they are the variation range relevant for the (crustal) scale of the problem we are dealing with.

By changing the focusing parameter ε, we can effectively control the level of sharpness of the result. For example, Fig. 6 shows the comparison of the results obtained with three different values of ε (but, of course, similar $\chi^2$ values; Fig. 7). With respect to the first inversion generated with $\varepsilon = 10^2$ (Fig. 1c and Fig. 6b), as expected, a smaller value of ε ($\varepsilon = 5$, in Fig. 6c) produces more numerous abrupt changes characterized by smaller mutual velocity differences (with $\varepsilon = 5$, the velocity jumps are of the order of tens of m/s); if, instead, we increase ε ($\varepsilon = 10^3$, in Fig. 6a), we can get a smoothly varying vertical profile since only the variations larger than the threshold defined by the focusing parameter will be penalized significantly. This shows how the MGS regularization can be effectively used in a deterministic framework to generate a wide range of solutions characterized by different levels of sparsity by means of a simple modification of the value of the focusing parameter.

For this test, an extensive study of the sensitivity of the final inversion with respect to the starting/reference model has not been performed and a homogeneous velocity ($[\boldsymbol{\beta}_0]_k$ = 3.4 km/s, $\forall k$) has been always used. However, as both (MGS and MGN) definitions include a derivative operator, we do not expect big impacts of the velocity values per se. In addition, even if starting from a model with a shape close to the real one would be, for sure, beneficial (even just in terms of convergence speed), it seems (Fig. 1 and Fig. 6) that the proposed approach can effectively handle quite arbitrary (i.e. distant from the actual solution) starting models.

### 3.2 Near-surface example

The second synthetic example consists of the near-surface model MODX discussed in Xia et al. (1999), Cercato (2007), and Haney & Tsai (2017). As described in Haney & Tsai (2017), the data associated to this model were generated with 2% noise, but, differently from that study, here, we use a homogeneous parameterization with 0.4 m-thick layers. By considering relevant for this test simply the velocity variations larger than a few meters per second, the value of the focusing parameter ε for the MGS inversion has been chosen equal to 5 (corresponding to expected velocity variations of ~12 m/s over a 1 m thickness). The corresponding result is plotted in Fig. 8a, whereas the MGN result with comparable data fitting (Fig. 9) is showed in Fig 8b. Both results are very good in retrieving the overall gradually increasing velocity profile. Nevertheless, the MGS performs better in inferring the discontinuity at around 13 m depth and the velocity of the last layer. In addition, even if it is not capable to retrieve all the small velocity variations (at least at this data fitting level: $\chi^2 \sim 0.81$), the MGS model is more successful in following the little shallow velocity steps.

With respect to the crustal example, here, the DOIs inferred by the MGS and MGN solutions look quite different (Fig. 10). However, as before, this is mainly due to the fact that the DOI concept, despite being useful for its immediacy (critical, e.g., in case of large surveys), cannot entirely convey the complexity of the sensitivity dependence with depth. A proper interpretation of the DOI value should take into account also the velocity profile. In the specific case, it is clear, that, also for the MGS result, the sensitivity variation with depth is very limited below 15 m as, below that level, the velocity (and sensitivity) is almost constant. So, the 25 m difference between the MGN's and the MGS's DOIs is merely due to the choice of the DOI threshold value and it is not really meaningful per se. As before, an analysis based on the integral sensitivity (and not just the DOI) can, instead, lead to a more correct interpretation (Fig. 10).

### 3.3 Geotechnical field test

We further verify the proposed tunable sparsity approach on an experimental dataset collected over a site characterized by the following lithological sequence (Fig. 11d):

    i) 0.0 to 0.3 m – clay till;

    ii) 0.3 to 1.5-2.0 m – artificial soil;

    iii) 1.5-2.0 m to ~10 m – till;

    iv) deeper than approximately 10 m – bedrock.

The geological settings are typical of the shallow subsurface around Lund (Sweden). However, in this specific case, the first 1.5-2.0 m have been altered. In particular, the original unit between 0.3 and 1.5-2.0 m have been substituted with artificial soil, and, consequently, also the most superficial layer (0-0.3 m) has been reworked.

The proposed tunable stabilizer behaves as expected. Hence, as for the synthetic example at crustal scale, also for this experimental test, by decreasing the value of ε, we obtain shear-wave velocity reconstructions characterized by an increasing level of sparsity. Fig. 11 shows three different solutions resulting from the application of ε spanning a wide range of values (from 6 to 9 10$^{-2}$). The parameterization consists of layers with a homogeneous thickness of 0.05 m; consistently, the selected ε values result to be sensitive to velocity changes of the order of, respectively, 120 m/s, 12 m/s, and 1.8 m/s over a meter. All the results in Fig. 11 fit the data equally well as show in Fig. 12 (the corresponding $\chi^2$ values ranges between 0.45 and 0.50) and all of them capture the velocity inversion at 0.3 m and detect the top of the bedrock.

However, the two sparsest solutions correctly retrieve the thickness and location of the till layer. This is particularly true for the solution with ε equal 0.6 (Fig. 11b), which, in fact, is sensitive to the most reasonable velocity change interval for the scale of interest of this specific test.

Concerning the DOI estimation, all three results in Fig.11 provide approximately the same depth. It is interesting to see that, at depth (below the DOIs), the solution is influenced uniquely by the starting model and is no longer sensitive to the data. In this specific test, the DOI value can be considered a reliable estimation of the depth at which the results stop being data-driven as the integrated sensitivity does not drop abruptly at a sharp velocity change, but rather cross the 70 dB threshold smoothly (Fig. 13).

### 3.3.1   Choice of the starting model, inversion robustness, and reliability of the DOI assessment

We use the experimental data to further analyze the dependence of the retrieved solution with respect to the choice of the starting (and reference) model $\boldsymbol{\beta}_0$. It is worth recalling that a reliable – and easy to get – estimation of the shear-wave velocity at depth is clearly provided by the value of the phase velocity at low frequencies as it can be immediately obtained by looking at the dispersion curves (e.g., in Fig. 12). Hence, we inverted the experimental data by using exactly the same settings except for the choice of the starting/reference model. More precisely, Fig. 14 shows the results of the inversion performed with $\lambda = 5\ 10^{-7}$ and $\varepsilon = 0.6$ (as for Fig. 11b) and homogeneous starting velocities equal, respectively, to 400 m/s, 500 m/s, and 600 m/s. Clearly, also in this case, the three results correspond to very close data misfits ($\chi^2 \sim 0.5$; see Fig. 15). The almost perfect overlapping of the solutions above the DOIs confirms the effectiveness of the proposed algorithm: for a reasonable range of starting velocities, the solution is stable with respect to the choice of the inversion parameters.

Besides, this test proves, once more, also the validity of the suggested DOI assessment (and, in particular, the reasonability of the 70 dB threshold). In fact, the three solutions begin to be significantly different below the DOIs, where each of them is highly affected by the specific choice of the starting model and not, anymore, by the data. Actually, as discussed for example in Oldenburg and Li (1999), the DOI could be assessed by performing several inversions and checking where the solutions are mainly dependent on the starting model (and not on the data). This alternative approach for the estimation of the DOI, despite being much more computationally expensive as it involves multiple inversions, might have (similarly to the proposed analysis of the integrated sensitivity) the advantage of further highlighting, not only at which depth the solutions start not being data-driven, but, also, if there are shallower "blind spots" (where the sensitivity of the model to the data is lower).

## 4 CONCLUSIONS

In this paper, we apply, for the first time to the inversion of surface wave dispersion curves, a tunable regularization based on the minimum gradient support. In particular, we provide practical suggestions on how to select the free parameters involved in the definition of this kind of regularization and demonstrate the capabilities of this inversion scheme in retrieving solutions, equally compatible with the observations, but characterized by different sparsity levels. So, by selecting the appropriate focusing parameter value, we can effectively span the solution space.

We first test the proposed approach on two synthetic – publicly available – benchmark datasets; they are about very different scales of application: from crustal to near-surface studies.

Moreover, we verify the proposed strategy on an experimental dataset collected for geotechnical purposes on a well-characterized site and compare the results against the known lithology.

In addition, aiming at a fair comparison of the results of the proposed tunable sparsity regularization, we push further some of the conclusions in Haney & Tsai (2017) and discuss the effectiveness of a strategy for the assessment of the depth of investigation based on the integrated sensitivity. In this respect, we highlight possible drawbacks of direct interpretation of the depth of investigation value (or values, depending on the number of propagation modes actually inverted), especially when dealing with abrupt velocity changes.

**ACKNOWLEDGMENTS**

Thanks are due to the Editor, Frederik Simons, and the two Reviewers, Andrea Zunino and Matthew Haney, whose comments largely contributed to the improvement of the original version of the manuscript.

This work was partially supported by Regione Autonoma della Sardegna (Legge regionale 7 agosto 2007, n.7 "Promozione della Ricerca Scientifica e dell'Innovazione Tecnologica in Sardegna") through the "Programma di mobilità dei giovani ricercatori 2018"; by the project "GEO-CUBE" (RAS/FdS – CUP F72F20000250007) by the project "Kvalitetskontroll av markstabilisering genom seismisk mätning med optisk fiber" (VINNOVA - 2018-00641); and by the project "Kontroll av markstabilisering med optisk fiber" (SBUF - 13579).

The authors thank Roger Wisén from Lund University for his support in planning and collecting the field data. In addition, Giulio Vignoli and Jeniffer Barreto are grateful to LABMAST (University of Cagliari), and, particularly, to Antonio Cazzani and Sergio De Montis for their help.

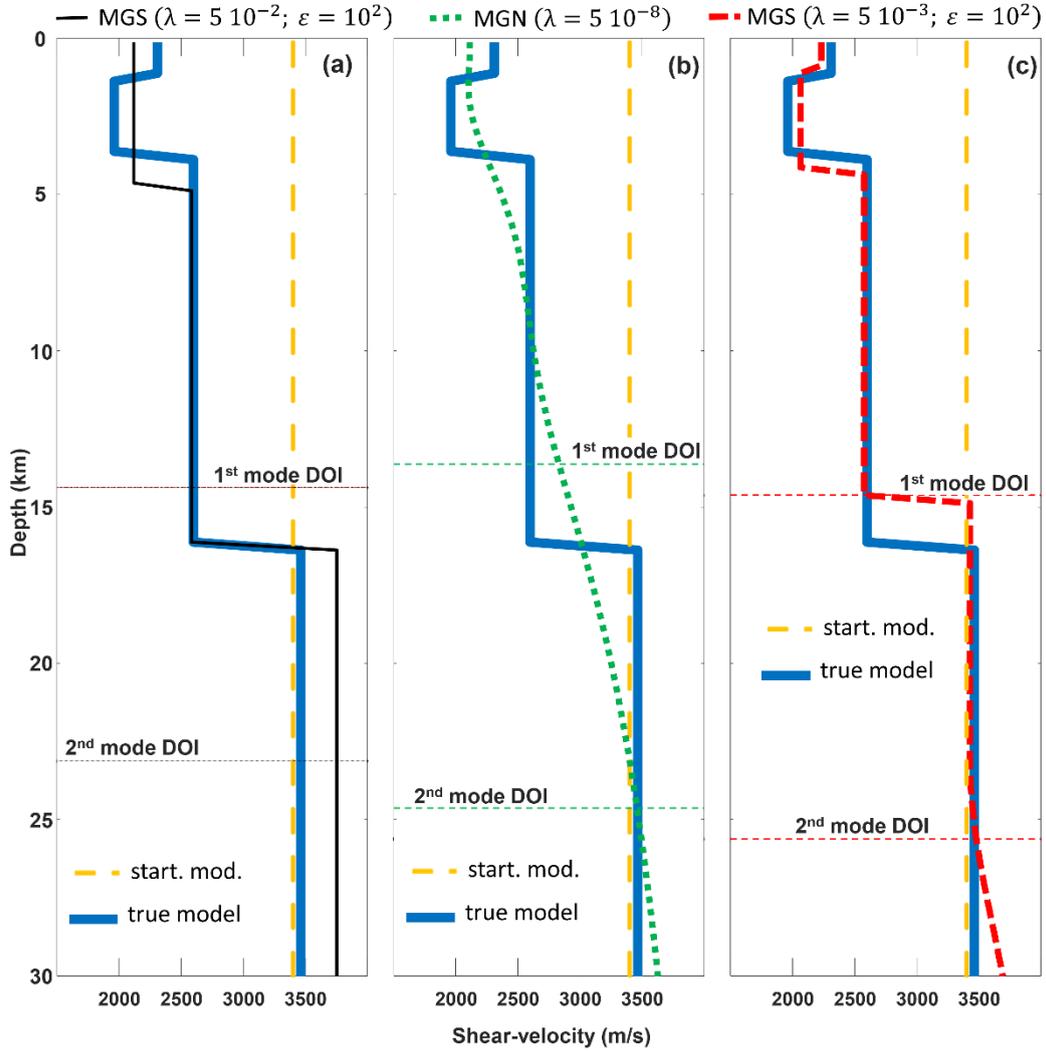

**Figure 1** Shear-wave velocity profiles related to the first numerical test (the crustal scale example). The solid thick blue line in all panels is the true model, whereas the dash orange line is the starting/reference model $\beta_0$. (a) shows the MGS reconstruction (solid black line) that can be consistently compared with the MGN result (dash green line) in (b) as they have compatible data misfit ($\chi^2 \sim 0.93$). In (c), the MGS result with a smaller $\lambda$ (and, coherently, with a lower $\chi^2$ value: $\sim 0.81$) is plotted (dash red line). For all three cases, the Depth Of Investigation (DOI) of each propagation mode is showed. All the inversion models consist of 240 layers of constant thickness (250 m) and the solutions are reached, respectively, in: 69 (a), 5 (b); 83 (c) iterations.

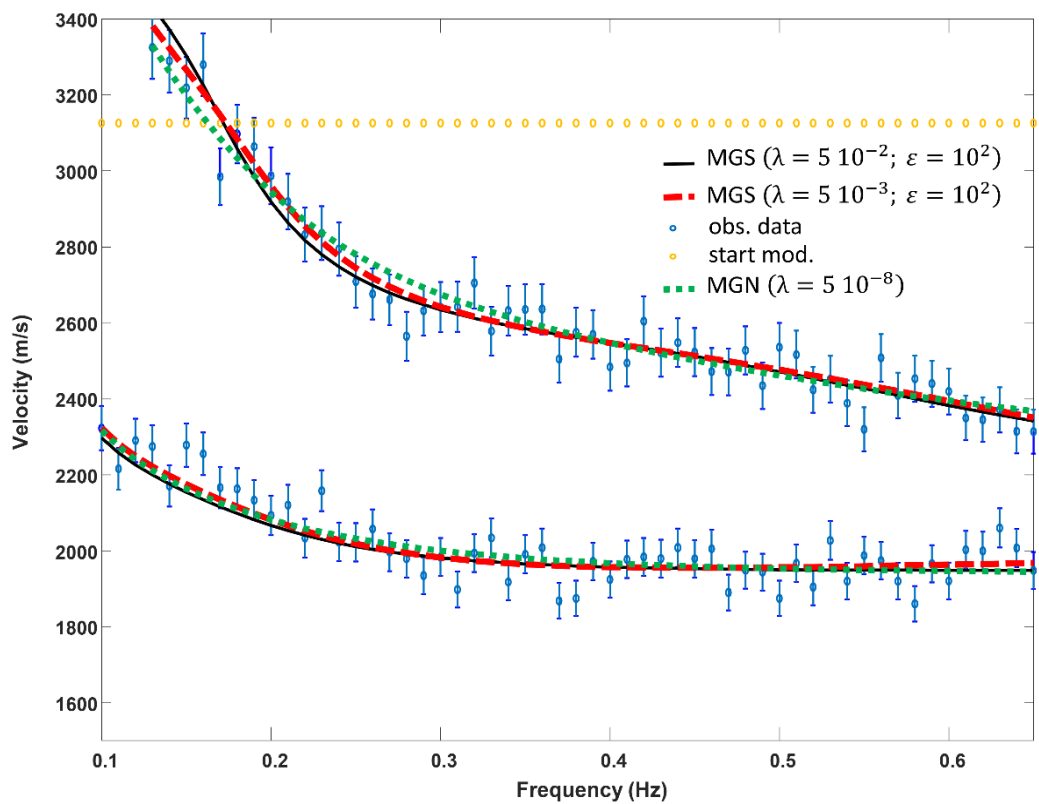

**Figure 2** The observed (blue circles with the associated error bars) and calculated data for the three inversion results in Fig. 1. The data from the MGS and MGN results with comparable $\chi^2$ values are plotted with a solid black and a dot green line, respectively, whereas the "overfitting" response obtained with the MGS (and a smaller $\lambda$) is represented by the dash red line. The orange circles are the data generated by the homogeneous starting model.

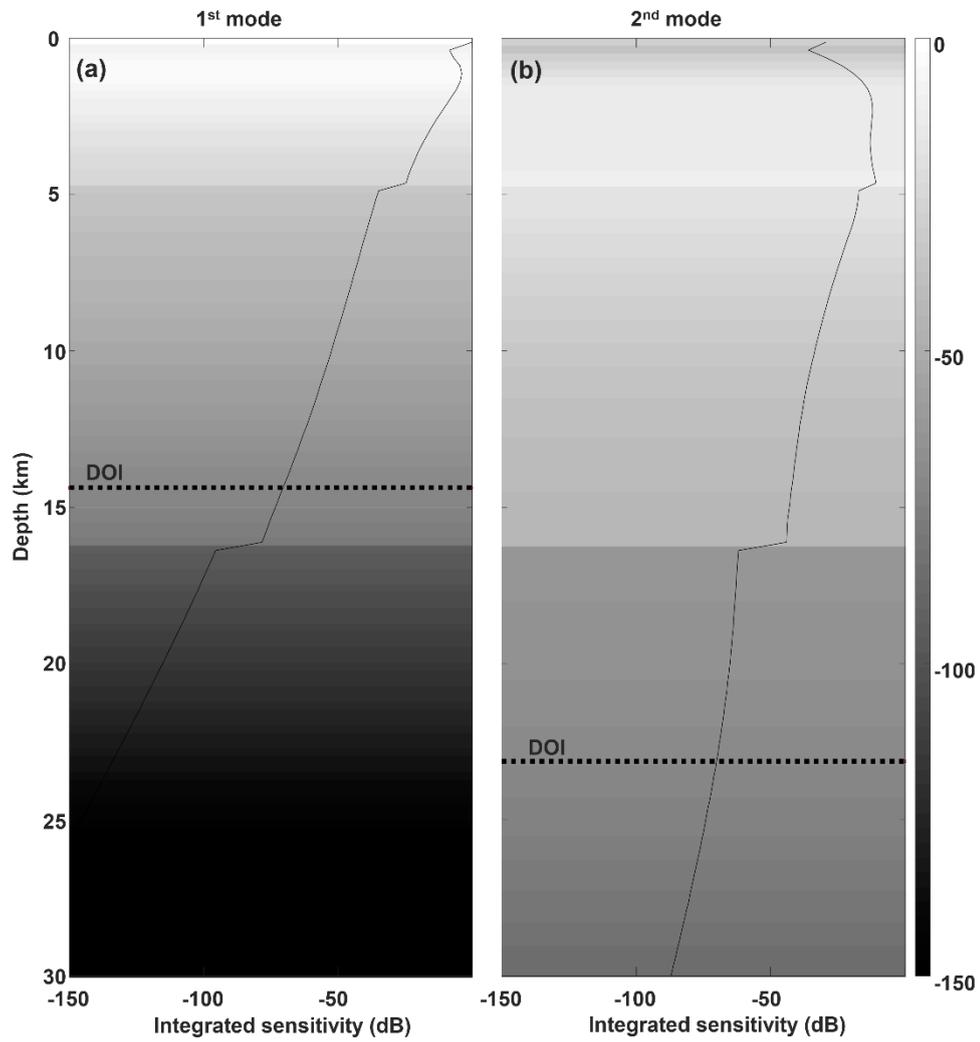

**Figure 3** Integrated sensitivity and estimation of the DOIs for the MGS ($\lambda = 5\ 10^{-2}$; $\varepsilon = 10^2$) solution in Fig. 1a. (a) and (b) show respectively the normalized integrated sensitivity for the fundamental mode and the first overtone. The DOI lines indicate at which depth the sensitivity drops by 70 dB.

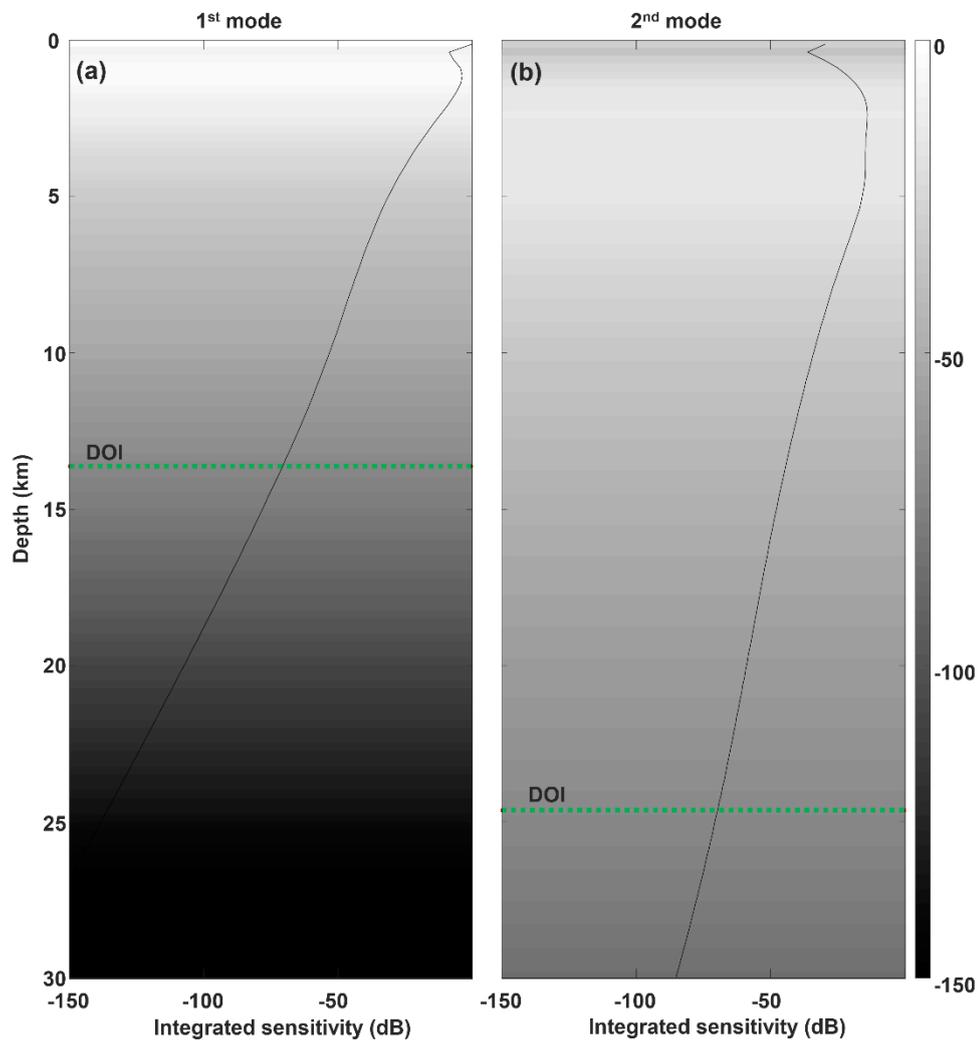

**Figure 4** Integrated sensitivity and estimation of the DOIs for the MGN solution in Fig. 1b. (a) and (b) show respectively the normalized integrated sensitivity for the fundamental mode and the first overtone. The DOI lines indicate at which depth the sensitivity drops by 70 dB.

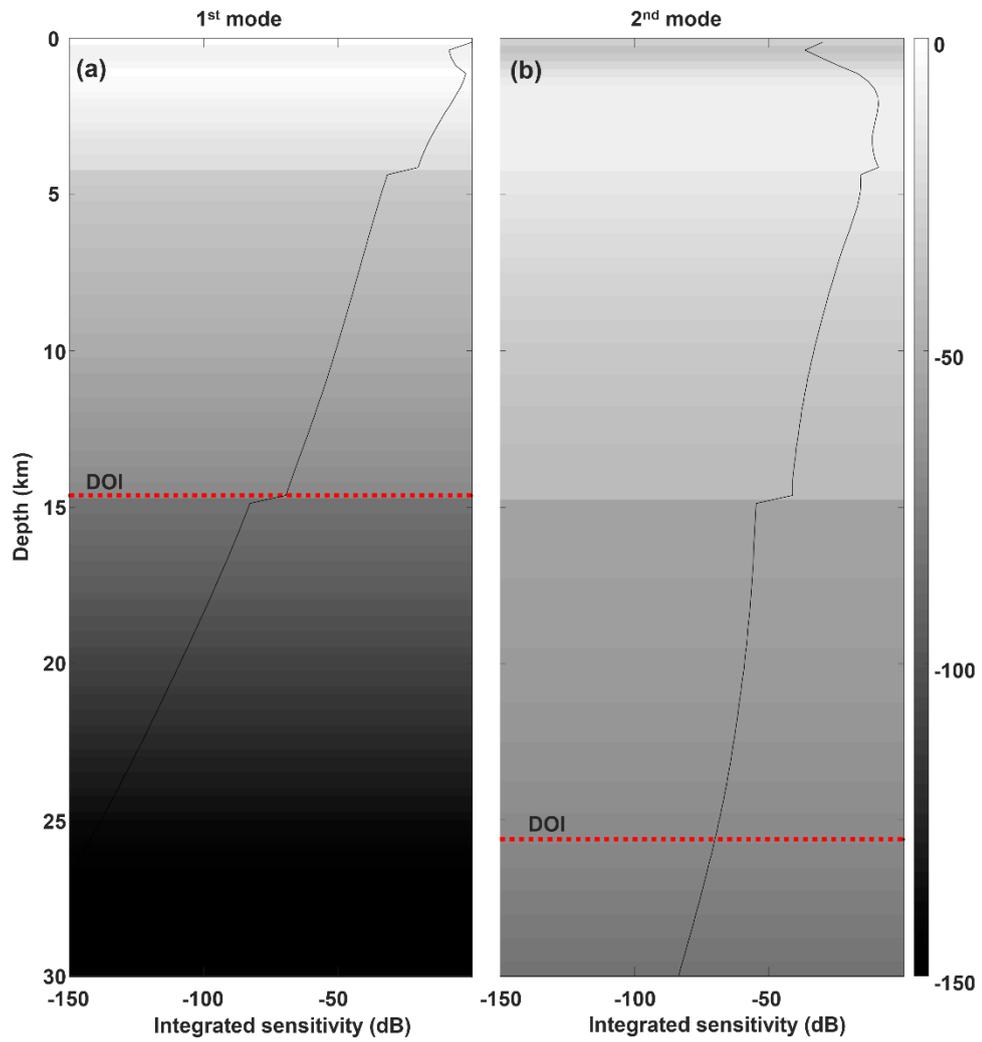

**Figure 5** Integrated sensitivity and estimation of the DOIs for the MGS ($\lambda = 5\ 10^{-3}$; $\varepsilon = 10^2$) solution in Fig. 1c. (a) and (b) show respectively the normalized integrated sensitivity for the fundamental mode and the first overtone. The DOI lines indicate at which depth the sensitivity drops by 70 dB.

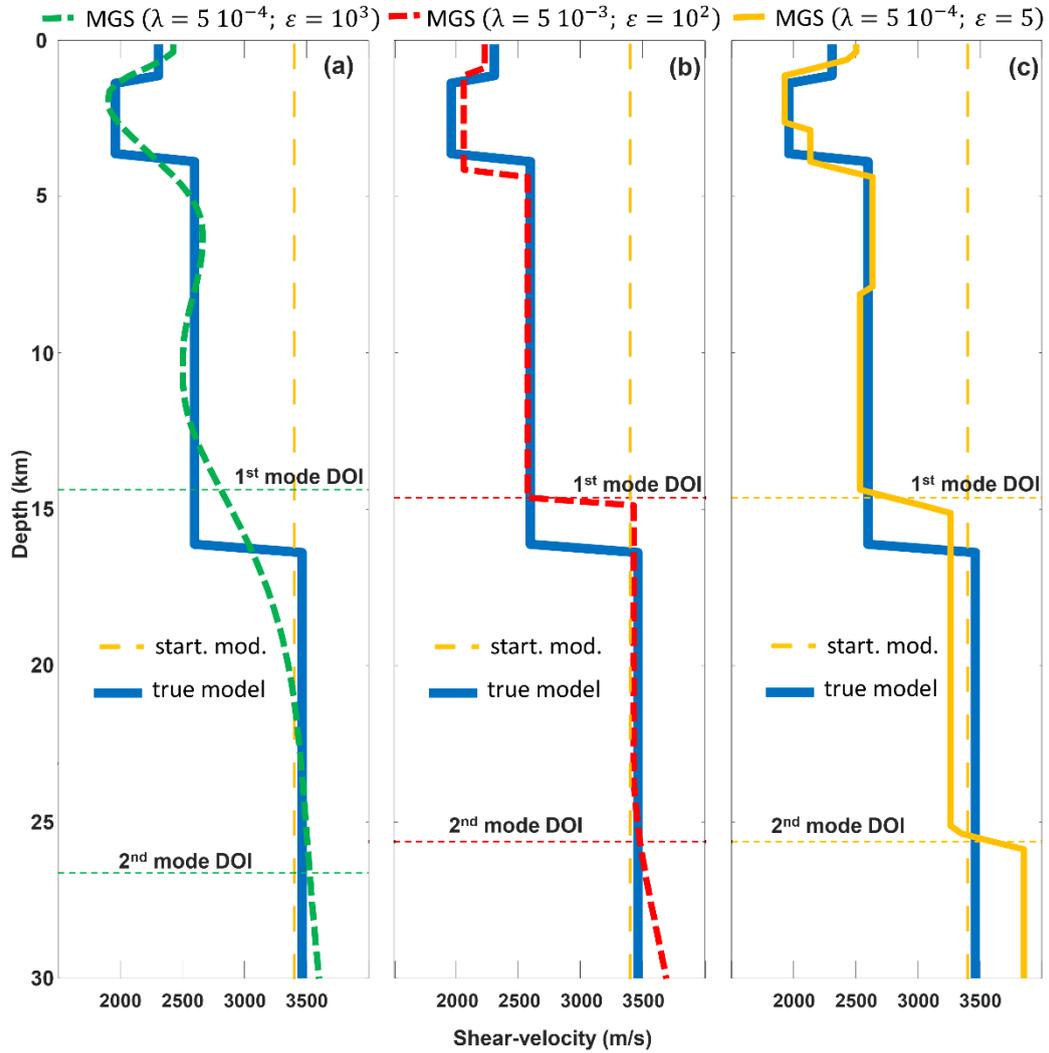

**Figure 6** Comparison of three MGS results for the first (crustal scale) example. Each of them is characterized by a different choice of the focusing parameter ε: (a) $\varepsilon = 1000$; (b) $\varepsilon = 100$; (c) $\varepsilon = 5$. All the inversion models consist of 240 layers of constant thickness (250 m) and the solutions are reached, respectively, in: 7 (a), 83 (b); 1572 (c) iterations. Of course, the data misfit for all three inversions is similar: $\chi^2 \sim 0.81$.

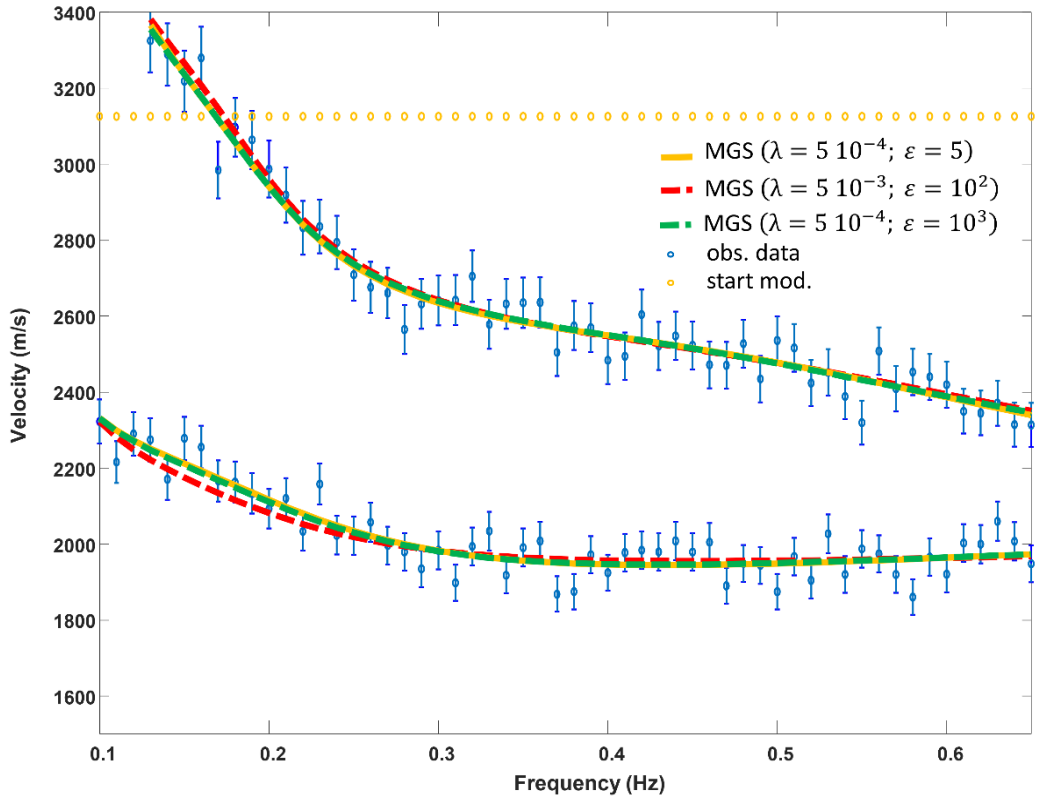

**Figure 7** The observed (blue circles with the associated error bars) and calculated data for the three inversion results in Fig. 6. The orange circles are the data generated by the homogeneous starting model.

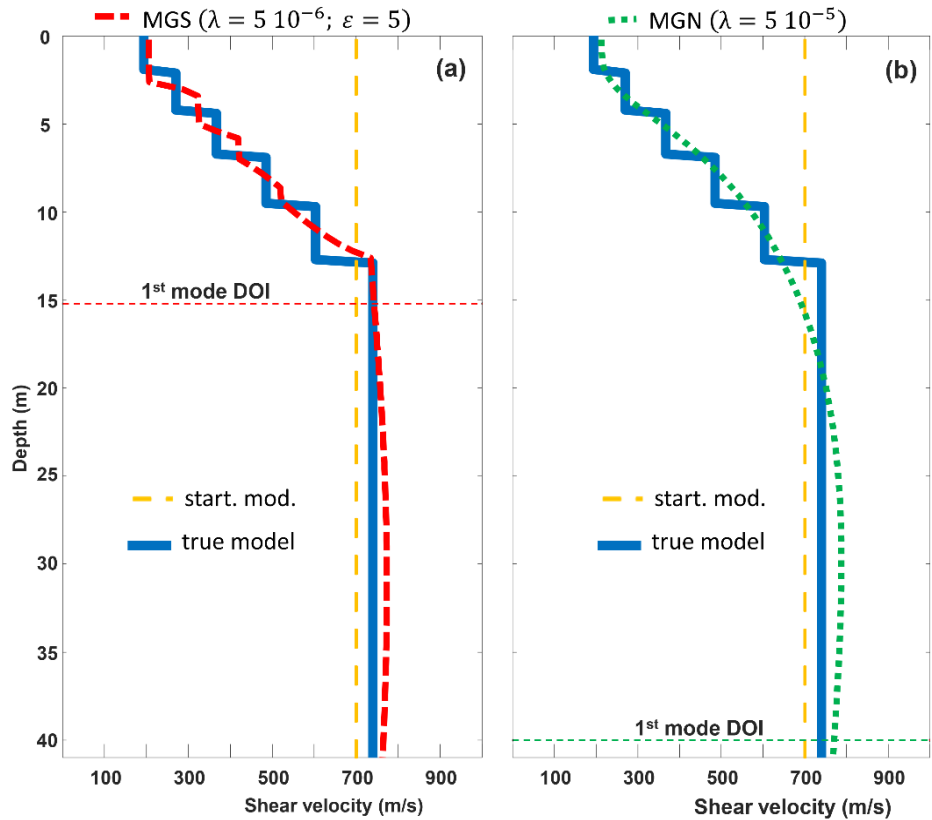

**Figure 8** Shear-wave velocity profiles related to the second numerical test (the near-surface example). The solid thick blue line in all panels is the true model, whereas the dash orange line is the starting/reference model $\boldsymbol{\beta_0}$. (a) shows the MGS reconstruction (dash red line) that can be consistently compared with the MGN result (dot green line) in (b) as they have compatible data misfit ($\chi^2 \sim 0.81$). For all two cases, the DOI for the fundamental propagation mode is showed. Both inversion models consist of 400 layers of constant thickness (0.4 m) and the solutions are reached, respectively, in: 228 (a) and 13 (b) iterations.

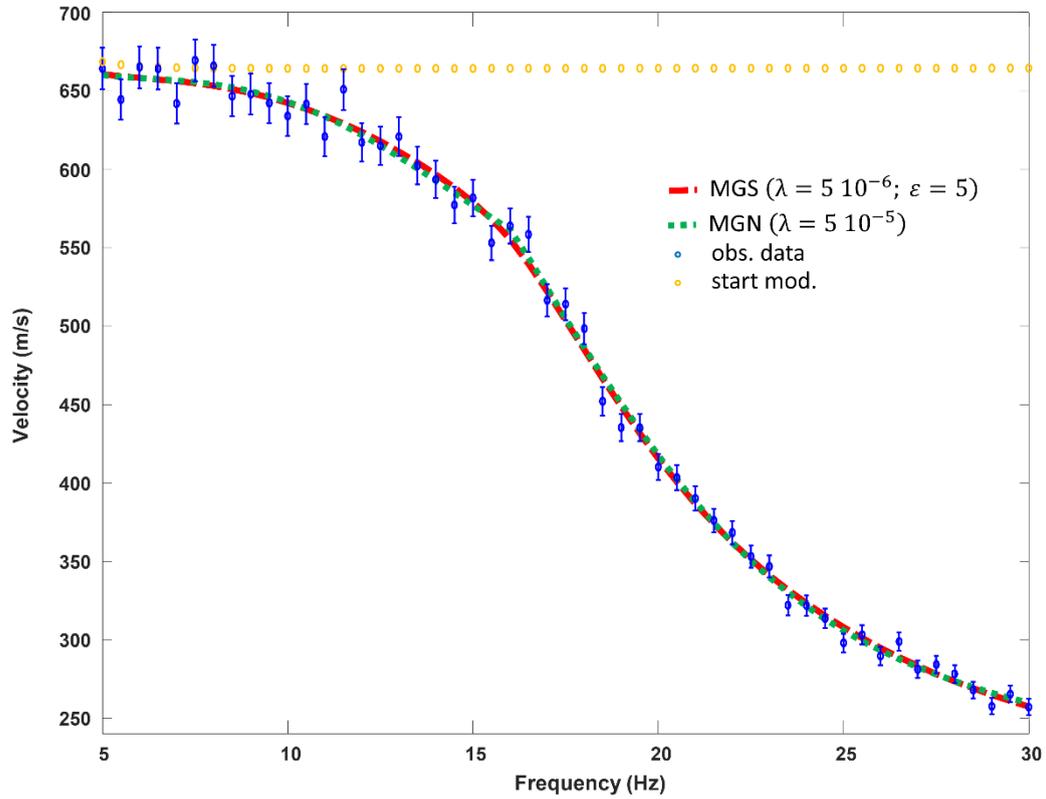

**Figure 9** The observed (blue circles with the associated error bars) and calculated data for the two inversion results in Fig. 8. The orange circles are the data generated by the homogeneous starting model.

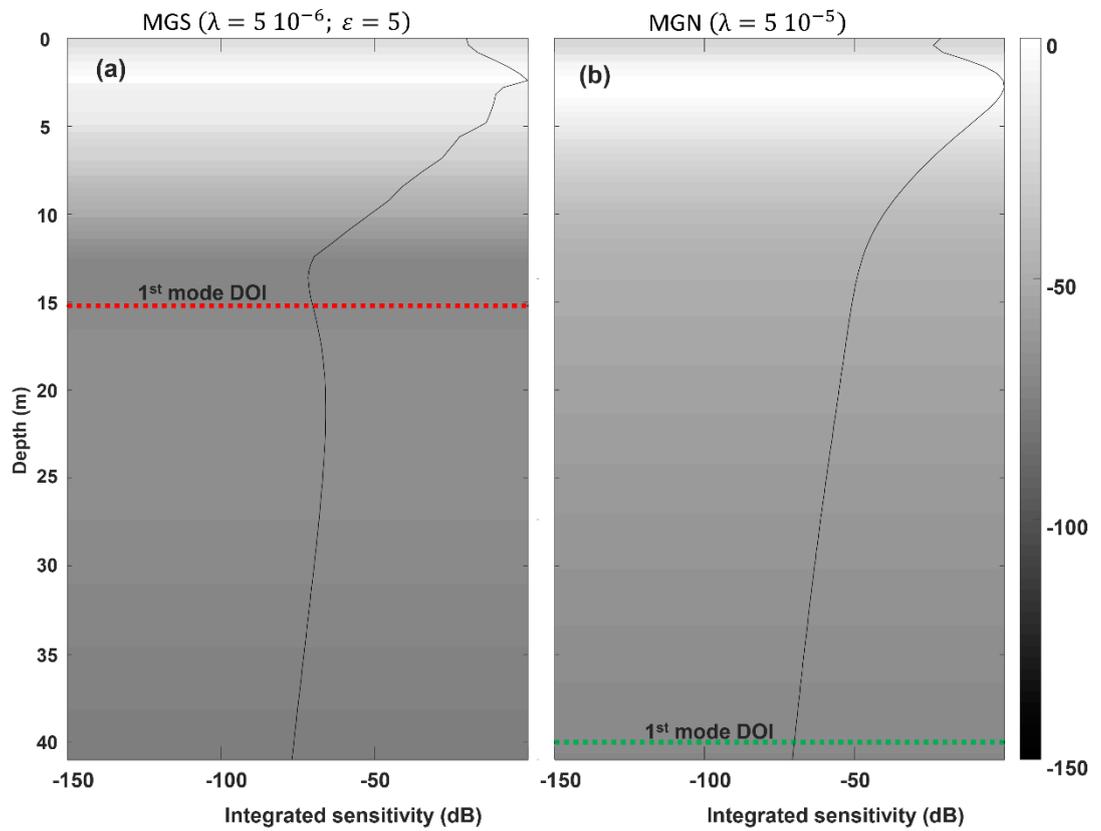

**Figure 10** Integrated sensitivity and estimation of the DOIs for the solutions in Fig. 8. (a) and (b) show respectively the normalized integrated sensitivity for the MGS and the MGN regularization. The two DOI lines indicate at which depth the sensitivity drops by 70 dB.

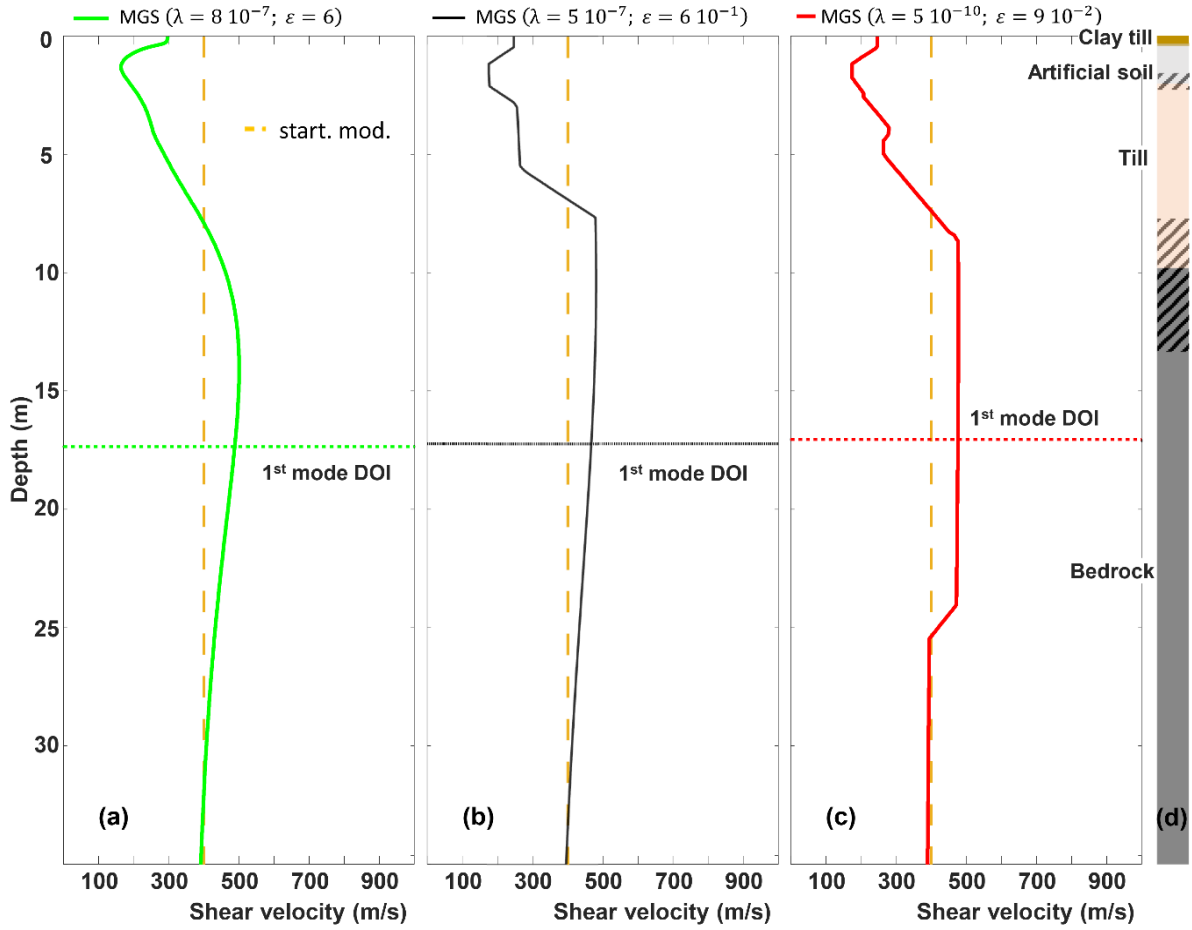

**Figure 11** Shear-wave velocity profiles retrieved from the data of the geotechnical experimental test. The green (a), black (b), and red (c) curves represent the solutions obtained with the MGS stabilizer and decreasing values of $\varepsilon$ (hence, increasing sparsity). All the solutions are characterized by $\chi^2$ values between 0.45 and 0.50. The dash orange line is the starting/reference model $\boldsymbol{\beta_0}$. In (d), the lithological sequence of the site is also shown (the dashed areas between 1.5-2.0 m and around 10 m represent the uncertainty related to the bottom interface of the artificial soil layer and the top of the bedrock). All models consist of 1000 layers of constant thickness (0.05 m) and the solutions are reached, respectively, in: 18 (a), 13 (b), 1788 (c) iterations.

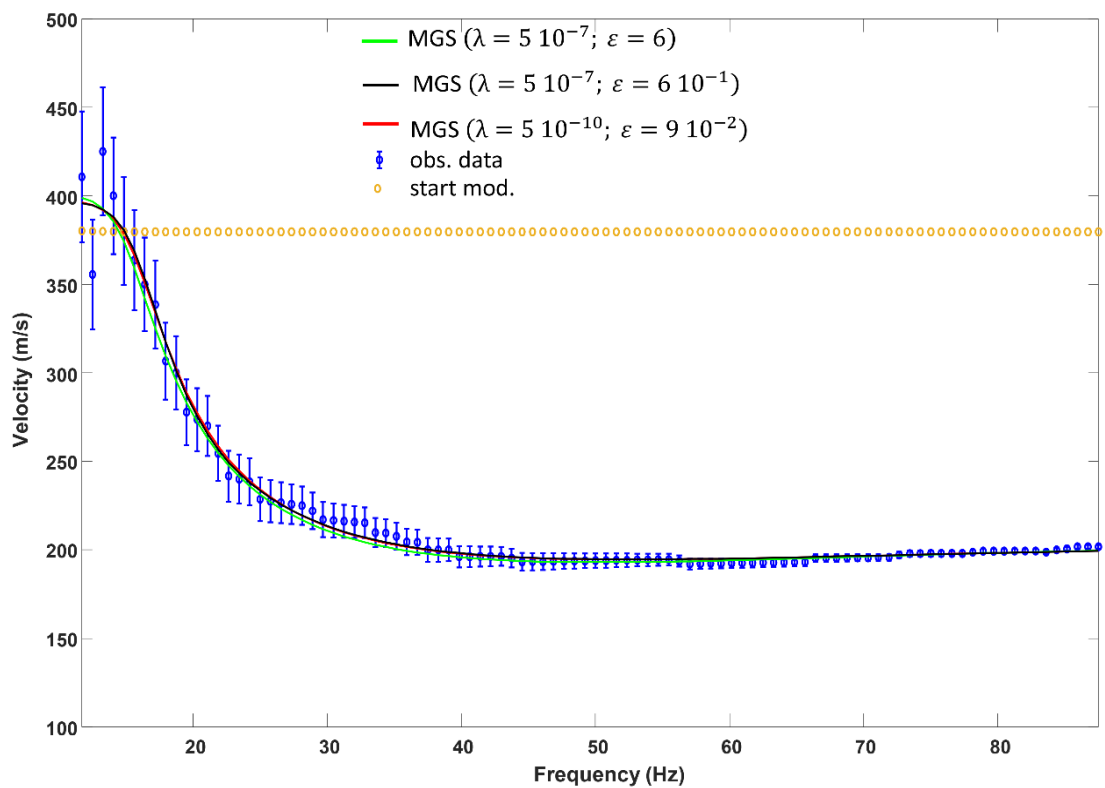

**Figure 12** The observed (blue circles with the associated error bars) and calculated data for the three inversion results in Fig. 11. The orange circles are the data generated by the homogeneous starting model.

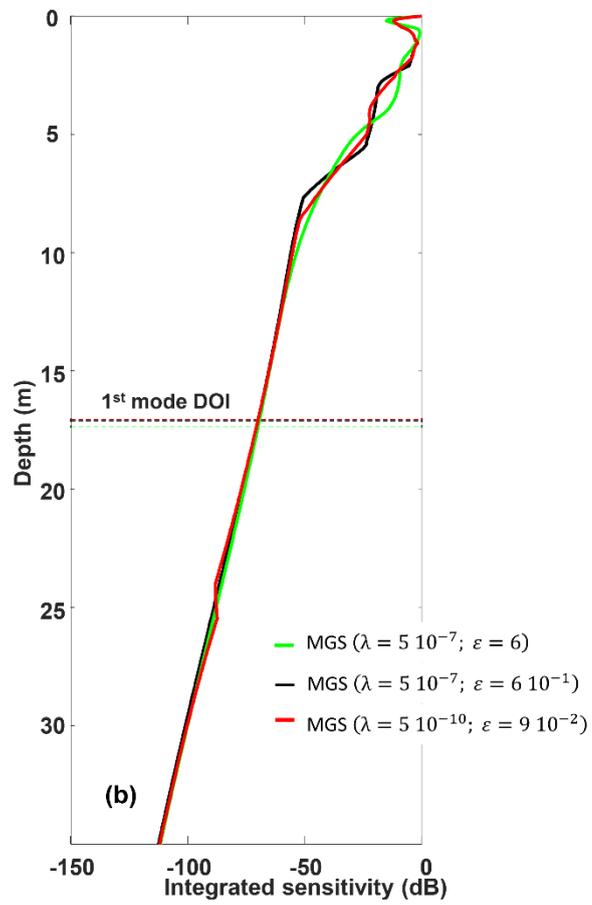

**Figure 13** Integrated sensitivity profiles and estimation of the DOIs for the solutions in Fig. 11.

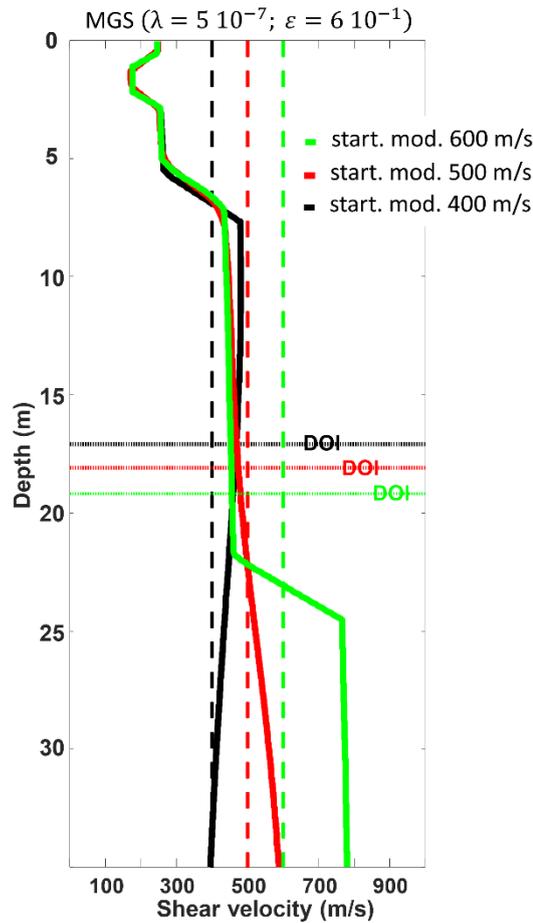

**Figure 14** Shear-wave velocity profiles retrieved from the data of the geotechnical experimental test. The green, red, and black curves represent the solutions obtained with the MGS stabilizer with the same settings used to obtain the result in Fig. 11b. The only difference between the three velocity profiles shown here is the starting/reference model. Together with the inversion results (solid lines), also the (homogeneous) starting models are shown (vertical dash lines) and the associated DOIs (horizontal dot lines). All the solutions are characterized by $\chi^2$ values about 0.5 (Fig. 15). All inversions are obtained with the same model parameterization consisting of 1000 layers (0.05 m thick); the solutions are reached, respectively, in: 75 (black), 51 (red), 95 (green) iterations.

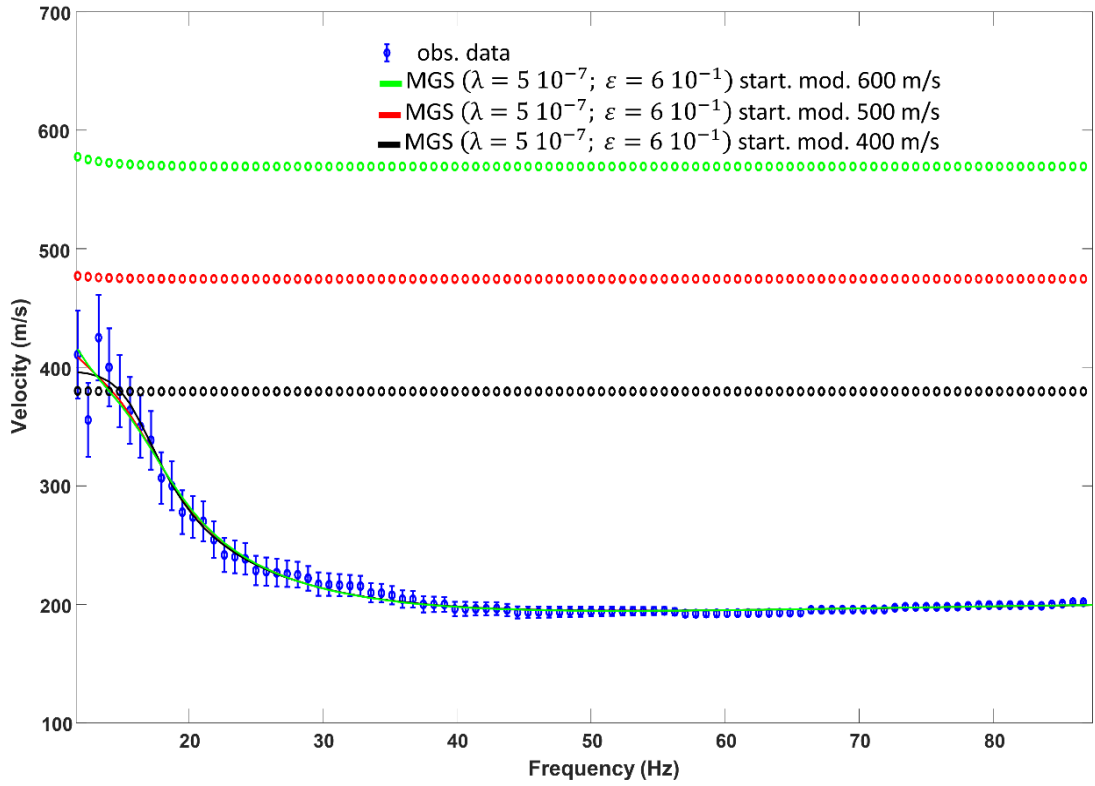

**Figure 15** The observed (blue circles with the associated error bars) and calculated data for the three inversion results in Fig. 14. The black, red, and green circles represent the data generated by the homogeneous starting models.